\documentclass[twocolumn, tighten]{aastex63}

\usepackage{amsmath,bm}
\received{ }
\revised{ }
\accepted{ }
\submitjournal{ApJ}

\shorttitle{Water in White Dwarf Systems}
\shortauthors{Becker et al. }

\begin{document}

\title{The Fate of Oceans on First-Generation Planets Orbiting White Dwarfs}

\correspondingauthor{Juliette Becker}
\email{juliette.becker@wisc.edu}

\author[0000-0002-7733-4522]{Juliette Becker}
\affiliation{Department of Astronomy, University of Wisconsin-Madison, 475 N. Charter Street, Madison, WI 53706, USA}

\author[0000-0001-7246-5438]{Andrew Vanderburg}
\affiliation{Department of Physics and Kavli Institute for Astrophysics and Space Research, Massachusetts Institute of Technology, Cambridge, MA 02139, USA}

\author[0000-0003-3888-3753]{Joseph R. Livesey}
\affiliation{Department of Astronomy, University of Wisconsin-Madison, 475 N. Charter Street, Madison, WI 53706, USA}

\begin{abstract}

Several groups have recently suggested that small planets orbiting very closely around white dwarf stars could be promising locations for life to arise, even after stellar death. There are still many uncertainties, however, regarding the existence and habitability of these worlds. Here, we consider the retention of water during post-main-sequence evolution of a Sun-like star, and during the subsequent migration of planets to the white dwarf's habitable zone. This inward migration is driven by dynamical mechanisms such as planet-planet interactions in packed systems, which can excite planets to high eccentricities, setting the initial conditions for tidal migration into short-period orbits. In order for water to persist on the surfaces of planets orbiting white dwarfs, the water must first survive the AGB phase of stellar evolution, then avoid being lost due to photoevaporation due to X-ray and extreme ultraviolet (XUV) radiation from the newly-formed white dwarf, and then finally survive the tidal migration of the planet inwards to the habitable zone. We find that while {this journey} will likely desiccate large swaths of post-main-sequence planetary systems, planets with substantial reservoirs of water may retain some surface water, especially if their migration occurs at later white dwarf cooling ages. Therefore, although stellar evolution may pose a challenge for the retention of water on exoplanet surfaces, it is possible for planets to retain surface oceans even as their host stars die and their orbits evolve. 

\end{abstract}


\section{Introduction}
\label{sec:intro}

Our Sun is destined to evolve into a white dwarf. This evolution will occur once the supply of hydrogen in the Sun's core runs out, at which point it will begin burning hydrogen in a shell around the now-inert core and expand into a red giant (RG) star, over 200 times its current size \citep{sackmann}. After there is no more fusable hydrogen remaining, the Sun will contract to about 20 times its current size and begin fusing helium in its core. Once the helium is in turn exhausted, the Sun will again expand to about 200 times its current size. But this time, instead of collapsing into another stable period of core fusion, the Sun will undergo several violent thermal pulses that expel its outer layers, leaving behind its exposed core that will cool into a white dwarf (WD). 

The Sun's evolution into a white dwarf will cause major changes to the the planets and smaller bodies in our Solar System as well. 
As the Sun ages, it will gradually become more luminous and make its planets warmer. Within about a billion years, the Sun's increased brightness will drive the Earth's climate into a runaway greenhouse state. 
The inner Solar System planets will be engulfed by the expanded star \citep{Rybicki2001, Schroder2008}, except perhaps for Mars, which may survive \citep{Veras2016b}. The surviving outer planets will change their orbits as the sun loses mass \citep{Duncan1998, Veras2012}. In addition to dynamical evolution of orbits, the remaining planets of the Solar System will be bathed with high levels of extreme ultraviolet radiation that can drive evaporation outflows in planets around Jupiter's orbital radius \citep{Villaver2007, Spiegel2012}. 
Dynamical instabilities brought about by this process possess the capability to significantly rearrange planetary orbits \citep{Voyatzis2013, Zink2020}. 
Such dynamical reshuffling is not exclusive to the Solar System; it will be seen broadly across exoplanetary systems as their host stars transition off of the main sequence (MS) stage \citep{Debes2002, Veras2015, Mustill2018, Stock2022, Kane2023}. 
Because of these processes, the end state of a planetary system orbiting a white dwarf will generally look substantially different than it looked on the main sequence. 

Despite the dramatic dynamical upheaval in planetary orbits will occur as a star evolves off the main sequence and becomes a white dwarf, planets in orbit around white dwarfs provide interesting prospects for habitability. {Planets that may not have been habitable while their host star was on the main sequence may, due to system reorganization spurred by stellar evolution, become habitable at later times. If that does happen, then t}he characteristic long cooling times of white dwarfs can potentially endow orbiting planets with extended habitable lifetimes. In fact, planets could conceivably reside in habitable orbits for up to approximately 3-10 billion years during this post-main-sequence phase \citep{Agol2011, Becker2023, Whyte2024}. Moreover, there is considerable evidence to suggest that planetary material is routinely transported to orbital radii close to the white dwarf's habitable zone, and at least some of the planetary objects remain intact \citep[e.g.][]{Jura2003, Zuckerman2010, Vanderburg2015, Gansicke2019, Vanderburg2020, Farihi2022}. 
Were a planet to reside in the habitable zone of a white dwarf, it would be a particularly favorable target for spectroscopic biosignature searches due to the small size of the white dwarf and resultant better contrast ratios compared to main sequence stars \citep{Loeb2013, Kaltenegger2020}.  

However, discussion of the habitability of planets around white dwarfs requires knowledge of not just the instantaneous surface temperature of the planet, as considered in works such as \citet{Agol2011} and \citet{Becker2023}, but also the surface conditions and their amenability to life. Because of the destructive influence of the red giant phase (which will destroy planets out to 2-3 AU), planets that reside near the habitable zone of a white dwarf (0.01 - 0.1 AU) will have been transported inwards after the white dwarf has formed\footnote{It is important to note that in this work, we consider only `first generation' planets, or those who existed while the star resided on the main sequence. Planets that formed after the star left the main sequence take different evolutionary pathways than considered in this work. }. In particular, the presence of surface water may be endangered during the thermally pulsing asymptotic giant branch (AGB) phase (when the star's luminosity increased by several orders of magnitude, which will highly insolate planets at large orbital radii), again during the young white dwarf phase (when the XUV luminosity of the white dwarf is at its maximum), and then once again during the planetary migration process that delivers a planet into a final, near-habitable orbit (when the planet may be significantly heated due to tidal forces). Even after planet migration is complete, remnant planetary eccentricity may lead to complete desiccation of the planet through continued tidal heating \citep{Barnes2013}.

In this paper, we consider whether oceans can be retained on planets that survive the red giant phase and subsequently migrate inwards to habitable orbits around the remnant white dwarf. 
In Section \ref{sec:hothothot}, we construct a simple model for planetary oceanic and atmospheric evaporation during the red giant phase and early life of the white dwarf, when its luminosity is high and it provides substantial irradiation even to distant surviving planets. 
In Section \ref{sec:migration}, we use a coupled model of orbital evolution and tidal heating to assess the dynamical evolution experienced by a planet after a strong dynamical perturbation that leaves a planet in a high-$e$ orbit. 
In Section \ref{sec:results}, we combine the models of the previous sections to demonstrate how a planet's initial orbital and physical parameters may shape the amount of water (if any) it is able to retain when it attains a habitable orbit. 
Finally, in Section \ref{sec:discussion}, we discuss extensions to this work, including ways to resupply oceans even once they are lost. 
 
\begin{figure*}
\centering
 \includegraphics[width=\textwidth]{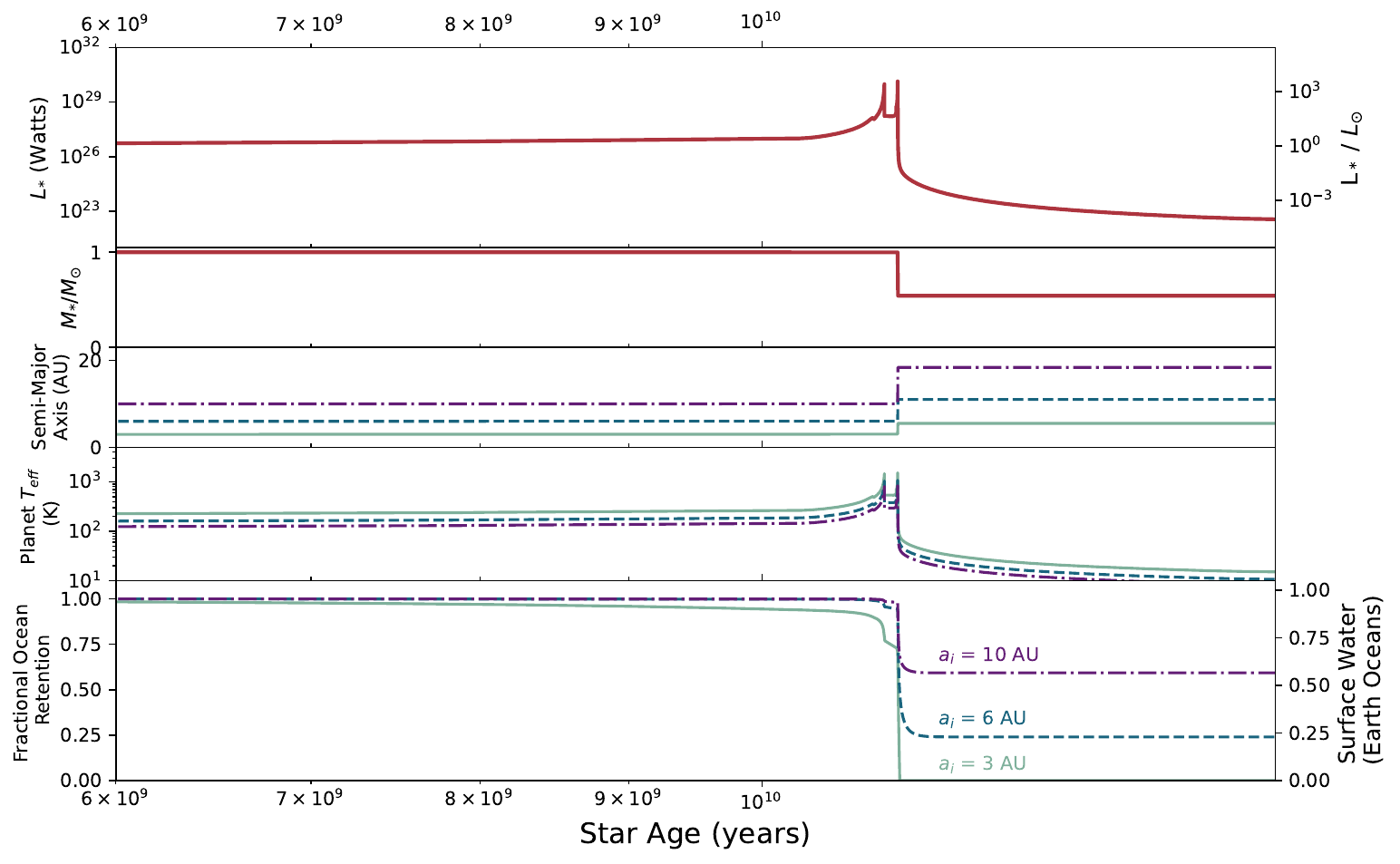}
 \caption{A model of the evolving properties of a Sun-like star as it evolves off the main sequence and eventually becomes a white dwarf, along with the resultant evolution of affected planet properties for a selection of initial planetary orbits. \emph{Top panel:} The luminosity evolution of the central body, constructed by combining the models of \citet{Hidalgo2018}, \citet{Bertelli2008}, and \citet{Salaris2022}. \emph{Second panel:} The mass of the central body \citep{Bertelli2008}. \emph{Third panel:} The semi-major axis of three planets with starting locations 3, 6, and 10 AU. \emph{Fourth panel:} The effective temperature of each planet based on its orbital location and the luminosity of the central body. \emph{Bottom panel:} The fraction of surface water remaining for an Earth-like planet with an initial 1 Earth ocean's mass worth of water. }
    \label{fig:lum_func}
\end{figure*}

\section{Surface Ocean Evaporation while the central body is hot}
\label{sec:hothothot}
As a low mass ($\leq 8 M_{\odot}$) star leaves the main sequence, its luminosity changes by several of orders of magnitude. 
The most substantial brightening occurs after the horizontal branch during the AGB phase, which is also when the star loses a substantial fraction of its mass. The luminosity evolution of a 1 $M_{\odot}$ star through this evolution is shown in the top panel of Figure \ref{fig:lum_func} (this luminosity evolution time-series is constructed from a combination of numerical models for a solar metallicity, solar-mass star: the MS/RG model of \citealt{Hidalgo2018}, the HB/AGB models of \citealt{Bertelli2008}, and the white dwarf cooling model of \citealt{Salaris2022})\footnote{The only correction performed in combining the models was to add a time offset to each model, matching $T_{eff}(t)$. The models of \citet{Miller2016} and \citet{Bertelli2009} were also used for comparison and verification purposes. }.

During the thermally pulsing asymptotic giant phase (TP-AGB), the pulsing star expels its outer shell, losing a substantial fraction (30-80\%; \citealt{Cummings2018}) of its mass and leaving behind a hot core which will become a white dwarf. 
Planets residing in the interior region of such a system will experience either physical engulfment \citep{Kunitomo2011, Zink2020} or tidal disruption during this time \citep{Nordhaus2013, Guidarelli2022}, depending on their orbital distance. 
However, even planets positioned at sufficiently large orbital distances to avoid direct engulfment or disruption will still be significantly impacted by their star's evolution \citep[e.g.,][]{Veras2017a, Maldonado2022}. During the relatively brief TP-AGB phase, such planets' orbits will expand, a process which is accompanied by drastic fluctuations in the stellar radiation the planets receive, varying by orders of magnitude. In this section, we explore how these aspects of stellar evolution influence the surface conditions and atmospheric compositions of planets that endure through this violent phase of a Sun-like star's post-main-sequence evolution. 

\subsection{Expansion of the Planetary Orbit}
Compared to the full stellar evolution, the stellar mass loss occurs over a relatively short amount of time: the $10^5 -10^6$ years of the AGB phase. 
The range of mass lost during this phase varies depending on stellar properties \citep{Bowen1988, Bloecker1995}, but for Sun-like stars the final white dwarf tends to end up being about half the mass of its progenitor \citep{Cummings2018}. Planets close to the red giant's radius will be affected by tidal forces resulting in more complex evolutionary pathways \citep[e.g.,][]{Guidarelli2022}. The evolution of more distant planets will be more simple: the orbits of planets residing significantly exterior to the red giant's physical radius will expand, {to} approximately conserve their angular momentum (see the third panel of Figure \ref{fig:lum_func}). 
These planets' expanding orbits reduce their exposure to the star's radiative flux. However, this occurs concurrently with an evolution in the central body's luminosity, which also increases significantly as the star approaches the tip of the AGB branch. 

The planetary effective temperature can be computed in the standard way using the planetary flux $F_{p}$, which is computed using the instantaneous values of the planet's orbital radius $a$ and the stellar luminosity $L_{\star}$:
\begin{equation}
F_p = \sigma T^4 = \frac{(1-A) L_{\star}}{16 \pi a^2}
\label{eq:radiation}
\end{equation}
where $T$ is the planetary temperature, 
$\sigma$ is the Stefan--Boltzmann constant and $A$ the planetary albedo (we assume Earth-like planetary albedo $A=0.3$ for the planets considered in this work). 
As the planet semi-major axis and central body luminosity $L_{\star}$ both evolve with time, this planetary temperature, while approximately constant while the host star is on the main sequence,  will also change with time post-main-sequence (see the fourth panel of Figure \ref{fig:lum_func}).

\subsection{A Model for Ocean Evaporation due to Irradiation}
Significantly exterior to the red giant's radius, planets will lose mass primarily via photoevaporative mass loss.
For a planet with a surface ocean, ocean loss is a multi-stage process. First, high surface temperatures must evaporate (for a water ocean) or sublimate/melt and evaporate (for an ice shell) the ocean into the atmosphere. 
Then, the water vapor must be dissociated into hydrogen and oxygen by incident high-energy photons. Finally, these atoms must escape into space at a rate that prevents later re-condensation as the white dwarf's luminosity decreases with age.

{Planets forming at 3-10 AU face unique formation conditions compared to terrestrial planets forming closer to their stars. Unlike terrestrial planets like Earth in the inner solar system, which may require volatile delivery from bodies formed in the outer Solar System to explain its water content \citep{Wanke1981,Hartogh2011,Taylor2024, Nimmo2024}, planets forming outside the snow line can naturally accumulate substantial water ices during their formation. The icy moon Europa illustrates this phenomenon. Europa’s ice shell is estimated to be 15-25 km thick, potentially holding twice the water volume of Earth's oceans \citep{Carr1998, Ojakangas1989, Pappalardo1999}, while other icy bodies in the Solar System like Callisto and Mimas also demonstrate the significant water reserves possible at similar or greater orbital distances \citep{Iess2014, Choblet2017, Lainey2024}. These observations support the plausibility of forming water/ice-rich bodies at 3-10 AU.}

To model the radiation-driven ocean evaporation and atmosphere escape, we follow the notation and procedure of \citet{Luger2015} and \citet{Barnes2020}. Note that in this approximation, the atmosphere is assumed to be initially fully hydrogen and low in mass compared to the ocean. As the planetary ocean evaporates, the atmosphere becomes a mix of hydrogen and oxygen, with a particle number density ratio of 2:1.

First, we can write the atmospheric mass escape rate for energy-limited flow driven by XUV radiation from the host star \citep{Watson1981,Erkaev2007,Lopez2012, Luger2015, Bourrier2017, Barnes2020} as
\begin{equation}
\dot{m}_{p,EL} = \epsilon \left(\frac{{r_p}^{3} L_{XUV}}{4 K  a^2 G m_{p}}\right),
\label{eq:mdot}
\end{equation}
where $\epsilon$ represents the heating efficiency, $r_p$ the planet's radius, $m_p$ the planet mass, $\rho$ its density, $L_{XUV}$ the stellar XUV luminosity, $G$ the gravitational constant, and $K = 1$ a tidal enhancement factor \citep{Erkaev2007} that approaches unity for long orbital periods.
The heating efficiency $\epsilon$ will vary as a function of the XUV flux between $\epsilon=0.1$ at early times and $\epsilon=10^{-3}$ or less at late times (as according to Figure 2 of \citealt{Bolmont2017}; see also \citealt{Yelle2004, Owen2013, Owen2016, Bolmont2017, Gallo2024}).

During the RG/AGB phases, we assume $L_\text{XUV}$/ $L_\text{bol} = 10^{-6}$ \citep{Schreiber2019}. Once the star has shed its envelope and become a WD, its XUV luminosity is a time-varying function of its temperature. 
We calculate $L_\text{XUV}$ by assuming the WD is an ideal blackbody; i.e., it radiates a Planck spectrum
\begin{equation}
u_\nu(\nu, T) = \frac{8\pi h \nu^3}{c^3} \frac{1}{e^{h \nu/k_B T} - 1},
\end{equation}
where $\nu$ is the frequency of the radiation, $h$ is the Planck constant, $k_B$ is the Boltzmann constant, and $c$ is the speed of light. We evaluate the fraction of the WD's bolometric luminosity that is emitted in the XUV waveband as
\begin{equation}
\frac{L_\text{XUV}}{L_\text{bol}} = \frac{\int_\text{XUV} d\nu \: u_\nu(\nu, T)}{\int_0^\infty d\nu \: u_\nu(\nu, T)}.
\end{equation}
For the purposes of this integral, the XUV waveband spans 1--1200 \r{A} ($2.498 \times 10^{15}$ Hz to $2.998 \times 10^{18}$ Hz). 
Assuming that all the mass lost is hydrogen atoms, this XUV flux will drive a particle escape flux of:
\begin{equation}
F_{H,ref} = \frac{\epsilon F_{XUV} r_p}{4 G m_p K m_H},
\end{equation}
where $F_{XUV}$ is the XUV flux received by the planet.

However, as the surface temperature of the planet grows hotter, then the ocean will fractionally evaporate, eventually leading to an atmosphere dominated by water vapor. In that case, we must also account for oxygen atoms that will be dragged along with the outflow \citep{Luger2015}, which will decrease the hydrogen particle flux. We can write the new particle loss rate: 
\begin{equation}
F_{H} = F_{H,ref} \left(1 + \frac{X_O}{1 - X_O} \frac{m_O}{m_H} \frac{m_c - m_O}{m_c - m_H} \right)^{-1},
\label{eq:mdot_steam}
\end{equation}
when $X_O$ is the oxygen molar mixing ratio \citep[which reaches a maximum of 1/3 at high temperatures where the atmosphere is pure water vapor]{Barnes2020}, $m_O$ is the mass of an oxygen atom, and $m_H$ is the mass of a hydrogen atom.
$m_c$ is the mass of the largest particle which can be driven upward by the hydrodynamic flow, which is defined as \citep{Hunten1987}:
\begin{equation}
    m_c = m_H + \frac{k_{B} T F_{H}}{b g X_H}
    \label{eq:mc}
\end{equation}
where $X_H$ is the hydrogen molar mixing ratio, $b = 4.8 \times 10^{17} (T$/ Kelvin$)^{3/4}$ cm$^{-1}$ s$^{-1}$ the binary diffusion coefficient \citep{Zahnle1986}, $g$ gravitational acceleration, $k_{B}$ the Boltzmann constant, and $T$ the temperature. Oxygen will by dragged along by the hydrodynamic flow and escape when $m_c > m_O$. As a comparison, \citet{Luger2015} finds that for an Earth-sized planet, oxygen will begin to escape once $F_{XUV} > 39 F_{\oplus}$, a value that is significantly exceeded during the AGB phase (top panel of Figure \ref{fig:lum_func}). When this occurs, the oxygen escape flux will be:
\begin{equation}
F_{O} = \eta F_{H} / 2
\end{equation}
where
\begin{equation}
\eta = \begin{cases}
0 \hspace{45mm} &x<1 \\
\left( \dfrac{x-1}{x+8}\right) \hspace{13mm} &x\ge1
\end{cases}
\end{equation}
where $x$ is determined by the mass of the largest particle which can be driven by the flow of hydrogen (Equation \ref{eq:mc}) and for oxygen is defined as $x = k_B T F_{H,ref} / (10 b g m_H)$. $\eta = 0$ corresponds to the case that only hydrogen escapes, while $\eta = 1$ corresponds to the case where entire units of water vapor (two hydrogen atoms and one oxygen atom) escape together (that is, all oxygen atoms will be dragged along with the escaping flow of H).  Then, the rate at which an ocean will be lost can be written as \citep[see Appendix A of][]{Luger2015}:
\begin{equation}
    \dot{m}_{ocean} = \dot{m}_{p,EL}  \left( \frac{9}{1 + 8 \eta} \right) .
    \label{eq:oceanloss}
\end{equation}
In the case $\eta = 0$, oxygen will not escape from the atmosphere, but due to the atmospheric escape of its associated hydrogen atoms it will not be able to re-condense into water, resulting in a ocean mass loss of 9 amu per hydrogen atom lost to space. In the case $\eta = 1$, all oxygen will be lost from the atmosphere as well as the ocean.  In the former case, where only the H component of the water vapor escapes, $O_2$ may either remain in the atmosphere \citep{Luger2015} or be absorbed by a surface sink \citep{Meadows2018}. Either way, the surface water will be lost.

\begin{figure}
\centering
 \includegraphics[width=0.5\textwidth]{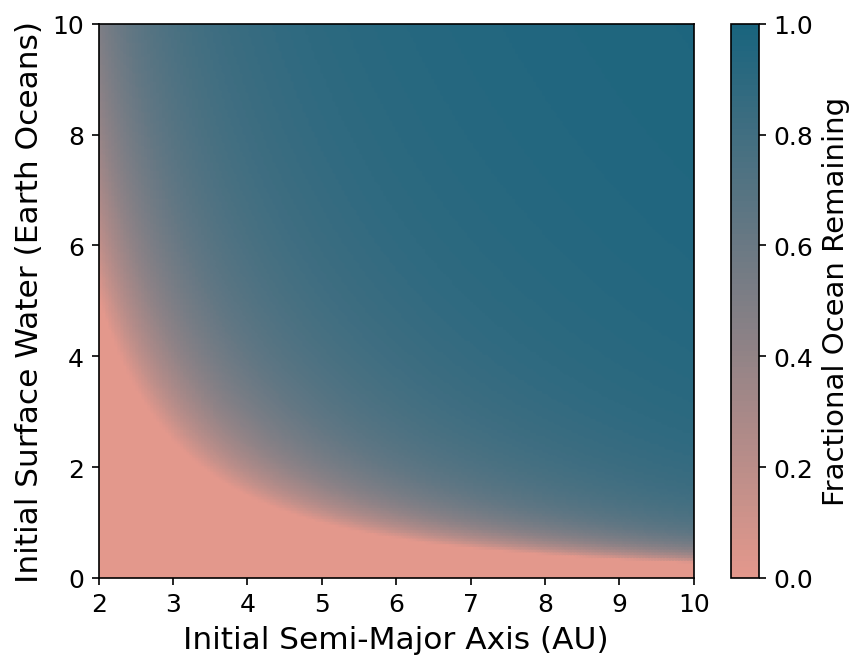}
 \caption{The fractional ocean remaining after 23 Gyr of main sequence and post-main-sequence evolution for planets with a variety of staring semi-major axes and initial ocean masses due to photoevaporative mass loss only. The calculation was performed as in the bottom panel of Figure \ref{fig:lum_func}, but only the final value of total surface water is reported. Surface water is not expected to change significantly past this time as the WD has cooled sufficiently to no longer drive outflows. Planets with Earth-like oceans are totally desiccated due to the radiative evolution of the host stars to fairly large initial orbital radii ($\sim5$ AU), but planets with larger initial oceans may retain significant amounts of water.   }
    \label{fig:ocean_left}
\end{figure}

In the bottom panel of Figure \ref{fig:lum_func}, we show the ocean retention fraction for three Earth-like planets that start with a surface water mass equal to 1 Terrestrial Ocean (TO, defined as the amount of water mass in Earth's present-day ocean), but with starting orbital radii of 3, 6, and 10 AU respectively. While planets at shorter orbital radii (as an example, the test case at 3 AU) may lose all their surface water before the AGB phase begins, more distant planets (those at 6 and 10 AU) will lose the largest amount of surface water during this highest-luminosity phase of stellar evolution. In these cases, the amount of surface water lost is primarily set by the duration of the AGB phase, as once the star concludes the TP-AGB phase and its remnant core starts to cool, further significant water loss due to XUV-driven processes becomes unlikely.

Some exoplanets may have oceans as deep as 2000 km \citep{Piaulet2023}, and in our own Solar System for several of the giant planets' satellites the amount of water on their surfaces exceeds the water budget of Earth \citep{Saur2015, Nimmo2016}. 
The census of exoplanets will include a large range of surface water masses, with planets past their systems' ice lines possibly harboring larger water reserves \citep{Bitsch2019}. While in Figure \ref{fig:lum_func} we assume that all our test case planets start with water mass budgets equivalent to that of Earth's ocean, there will in reality be a lot more variation in planetary water budgets. Figure \ref{fig:ocean_left} depicts the remaining ocean fractions for various initial conditions of semi-major axis and ocean mass. Planets with large initial ocean masses are more likely to retain them, particularly at larger semi-major axes where lower XUV flux allows a broader range of ocean mass fractions to preserve their surface water.

The evolution of a star off of the main sequence through the TP-AGB phase is highly destructive, both in terms of the dynamical reshuffling that can occur in the system and the surface changes driven by the heightened luminosity of the central body. 
Planets that were in the habitable zone while the star resided on the main sequence will likely lose all surface water, even if they survive the death of the host star \citep[which they usually will not;][]{Zink2020}. 
In contrast, more distant planets with water budgets that may have been frozen solid during most of the main sequence might retain this water once the star has completed its violent evolution and transitioned into a cooler white dwarf. However, such a planet's journey (and the journey of its surface water) is not finished once the violent phases of stellar evolution have concluded.

%


\section{The Model of Planet Migration and the Response of the Planetary Oceans}
\label{sec:migration}
For first-generation planets that end up near the habitable zone of the WD, significant orbital changes must occur. As the host sheds its envelope during the AGB phase, a planet's orbit will expand in radius. Then, to subsequently migrate to near the habitable zone (0.01 - 0.1 AU), a dynamically hot process must first excite the planet's orbital eccentricity and allow its periastron distance to reach roughly the habitable zone. Finally, tides raised on the planet's surface will circularize its orbit and bring it to its final (potentially habitable) orbital location, in a process analogous to tidal migration of hot Jupiters around main sequence stars. 

These significant changes to the planetary orbital parameters may also affect the water budget of the planet. In this section, we consider the planet's orbital evolution will alter the water content of the planet.

{\subsection{Feasibility of Planetary Scattering Events}
\label{sec:scatter}}

{The processes that can drive planets into high-eccentricity orbits and subsequently transport them inwards are varied and well-studied. Mechanisms include planet-planet scattering \citep{Veras2015}, perturbations by an unbound exterior star, Lidov-Kozai interactions with a bound companion \citep{Munoz2020}, and in some {orbital} geometries, the effects of stellar mass loss \citep{Adams2013b}. These dynamical changes have different timescales on which they can alter the orbits of planets and are spurred by stellar evolution, which alters the dynamical character of the system.} 

{For massive planets, multi-planet systems that are stable during the main sequence can become unstable via Hill or Lagrange instabilities post-main-sequence \citep{Veras2012}. Planet-planet interactions in such systems can increase orbital eccentricities, driving planets into white-dwarf-crossing or short-period orbits \citet{Veras2013}. For terrestrial planets, \citet{Veras2015} showed that instabilities in packed systems of smaller planets can result in some planets being tossed into high-eccentricity, short-pericenter orbits. Further, numerical simulations by \citet{Payne2017} showed that exomoons liberated from their orbits around planets during phases of strong dynamical evolution may also attain white dwarf-centric orbits with pericenters as small as 0.01 AU. While tidal evolution was not included in these previous models, tidal effects can provide the mechanism to subsequently circularize orbits.  }

{Notably, Hill instabilities in multi-planet systems can occur tens of millions to billions of years into the white dwarf cooling phase \citep{Veras2013, Mustill2014, Veras2015, Veras2016}. These delayed scattering events are distinct from continuous mechanisms like Lidov-Kozai oscillations, which would begin operating immediately after the post-main-sequence geometry is established. While the chaotic nature of these planet-planet interactions makes precise predictions of the rate of planet occurrence near white dwarf habitable zones challenging \citep[i.e.,][]{Veras2021}, the mechanisms to place planets in high-eccentricity orbits at a variety of white dwarf cooling age are feasible. }

\subsection{Thermal Effects of the Circularizing Orbit}
\label{sec:thermal}

While not all first-generation planets will migrate inwards near the habitable zone post-WD formation, those that will must be scattered inwards and start on very high eccentricity orbits. 

The location of the habitable zone around the white dwarf ranges between .001 and 0.1 AU, depending on the age of the white dwarf and what heating mechanisms are relevant (radiative versus tidal heating).
The effects of tidal heating on a planet's orbital parameters, given an initial non-zero orbital eccentricity, can be described using the following coupled differential equations \citep{Goldreich1963, Hut1981}:
\begin{equation}
\begin{split}
\frac{da}{dt} = -\sqrt{\frac{GM_\star^3}{a^{11}}}\,\bigg[ \frac{9}{2}\frac{\zeta\ k_{2,\star}}{Q_{\star}}  M_\star^{-2} R^5 m_{p} + \\
\frac{21\ k_{2,p}}{Q_{p} m_p}   \frac{r_p^5}{(1-e^2)^{15/2}} \biggl(1 + 31 \frac{e^2}{2} + 255 \frac{e^4}{8} + 185 \frac{e^6}{16} \\- (1-e^2)^{3/2} \left(1 + 15\frac{e^2}{2} + 45 \frac{e^4}{8} + 5 \frac{e^6}{16}  \right) \biggr)
   \bigg]\, 
\end{split}
\label{eq:dadt}
\end{equation}
and
\begin{equation}
\begin{split}
\frac{de}{dt} = - e\,\sqrt{\frac{GM_\star^3}{a^{13}}}\,\bigg[ \frac{171}{16}\frac{\zeta\ k_{2,\star}}{Q_{\star}} M_\star^{-2}  R^5 m_{p} +\\
\frac{21}{2}\frac{k_{2,p}}{Q_{p} m_p} \frac{r_p^5}{(1-e^2)^{13/2}} \biggl(1 + 15 \frac{e^2}{4} + 15 \frac{e^4}{8} + 5 \frac{e^6}{16} \\- \frac{11 (1-e^2)^{3/2}}{18} \left(1 + 3\frac{e^2}{2} + \frac{e^4}{8}\right) \biggr).
\end{split}
\label{eq:dedt}
\end{equation}
In Equations (\ref{eq:dadt}) and (\ref{eq:dedt}), $Q$ quantifies tidal dissipation efficiency due to distortions, represented by $Q_p$ for the planet and $Q_\star$ for the star \citep{Goldreich1966}. $k_{2}$ denotes the tidal Love number, expressing tidal force-induced deformation. $m_p$, $r_p$, $G$, $M_{\star}$, $R_{\star}$, $a$, and $e$ represent planetary mass, radius, gravitational constant, stellar mass, radius, planetary orbital semi-major axis, and orbital eccentricity, respectively. $\zeta = {\rm sign}(2\Omega_{\star} - 2n)$ signifies the direction of change of planetary parameters, dependent on relative frequencies of stellar spin rate ($\Omega_{\star}$) and planetary mean motion ($n$).

During this orbital evolution, the power dissipated by tidal strain $dE_p/dt$ can be written as (\citealt{Hut1981}):
\begin{equation}
\begin{split}
    \frac{dE_p}{dt} = \frac{21}{2} \frac{k_2}{Q_p} G^{3/2} M_{\star}^{5/2} r_p^{5} a^{-15/2} e^2 (1-e^2)^{-15/2}\\\times \left(1 + 15 e^2 / 4 + 15 e^{4} /8 + 5 e^{6} / 64\right)\,.
    \end{split}
    \label{eq:tidalheating}
\end{equation}
Given the substantial orbital eccentricities under consideration, we adopt the full eccentricity expansion from \citet{Hut1981} for all evolutionary expressions above, in lieu of the commonly used low-eccentricity expansion seen in exoplanet literature ($da/dt \propto 
e de/dt \propto dE/dt \propto e^{2}$).
If a planet is sufficiently deformable, this energy can be dissipated as heat in the planetary interior.
Planets with thin lithospheres and large mantle melt fractions will transport heat more efficiently, resulting in hotter surface temperatures. 

Here, by necessity, we must make some assumptions about how the energy is dissipated on the planet. In general, higher rates of energy dissipation will lead to higher temperatures on the planet surface, and subsequently higher rates of ocean evaporation and subsequent atmospheric mass loss, but the dissipation rate also sets the thickness of a planet's ice shell and ocean which in turn affects the dissipation rate. 
As such, the details of that calculation depend on the exact internal structure of the planet. For planets with molten cores, tidal heating may induce a runaway melting of the mantle, resulting in significant surface turnover, a mechanism that may be relevant for exoplanets orbiting main sequence stars \citep{Seligman2023}. However, the older planets orbiting white dwarfs have likely lost the heat due to radiogenic heating that drives core melt for young planets, suggesting that the initial condition for planets under our consideration is a solid core. Given our focus on ocean loss, we are considering the thermal and tidal evolution of planets with significant surface water/ice layers. In that case, heat will primarily be dissipated through the lithosphere.
For planets located originally beyond the ice line and possessing substantial surface water, their water is expected to exist as a subsurface ocean beneath a considerable layer of ice. This scenario mirrors the conditions observed on several moons in our solar system, such as Europa.
When these planets undergo migration, the heat produced by tidal forces can melt their ice layers. This process alters the planet's tidal quality factor over time. While a detailed analysis of how this quality factor evolves during migration is beyond the scope of this work, it is important to note that changes in the tidal quality factor will affect the energy dissipation and the effective temperature of the migrating planet.

In general, the surface temperature of the exoplanet during migration should be such that the tidal heat production inside the system can be radiated away. Meanwhile, the tidal dissipation depends on the temperature and the thickness of lithosphere, which is in turn affected by the surface temperature and the heat production. To solve this coupled system, one can follow the procedure taken by \citet{Ojakangas-Stevenson-1989:thermal} to estimate the equilibrium ice shell thickness and surface heat flux for a tidally-heated icy satellite, Europa. We assume that tidally generated heat is mostly produced in a conductive lithosphere. In thermal equilibrium, heat conduction should carry away the heat produced due to tidal dissipation $q$,
\begin{equation}
    \frac{d}{d z}\left(\kappa \frac{d T}{d z}\right)=-q,
    \label{eq:heat-conduction}
\end{equation}
where $\kappa=\kappa_0/T$ is the heat conductivity \citep{birch1940thermal}. Multiplying $\frac{d\ln T}{dz}$ to both sides of Eq.\ref{eq:heat-conduction} yields
\begin{equation}
    \frac{\kappa_0}{2}\frac{d}{d z}\left( \frac{d \ln T}{d z}\right)^2=-q \frac{d\ln T}{dz},
    \label{eq:heat-conduction2}
\end{equation}
and then integrating from the bottom of the lithosphere (which has a temperature of the melting point $T_m$) of the top of the lithosphere (with a cooler surface temperature $T_s$) yields:
\begin{equation}
    \frac{\kappa_0}{2}\left.\left( \frac{d \ln T}{d z}\right)^2\right|_{\mathrm{bot}}^{\mathrm{surf}}=\int_{T_s}^{T_m}\frac{q(T)}{T}dT
    \label{eq:heat-flux-surf-bottom}
\end{equation}
Such integration is more tractable because heat production at different depth only varies with the material mechanical property, which is a function of $T$, given the fact that all layers in the thin lithosphere undergoes almost the same deformation.
Assuming Maxwell rheology, the dissipation rate $q\equiv\overline{\sigma_{i j} \dot{\varepsilon}_{i j}}$ ($\sigma_{ij}$ denotes stress tensor and $\dot{\varepsilon}_{i j}$ denotes strain rate) can be estimated as
\begin{equation}
    q(T)=\frac{2 \mu \overline{\dot{\varepsilon}_{i j}^{2}}}{\omega}\left[\frac{\omega \tau_{\mathrm{M}}}{1+\left(\omega \tau_{\mathrm{M}}\right)^{2}}\right],
\end{equation}
where $\tau_M=\eta(T)/\mu$ is the Maxwell time, $\eta$ is the material's viscosity, $\mu$ is the media's rigidness, $\omega = 2 \pi / P$ is the frequency of the external forcing, and $P$ is the planetary orbital period. 
If we assume the viscosity varies with temperature according to $\eta(T)=\eta_m(T/T_m)^{-l}$, the integral on the right-hand-side of Eq.\ref{eq:heat-flux-surf-bottom} can be estimated as follows:
\begin{equation}
\begin{split}
\int_{T_s}^{T_{m}} q(T) \frac{d T}{T}=
\frac{2 \mu}{\omega l} \overline{\dot{\varepsilon}_{i j}^{2}}(\phi, \lambda) \huge[\tan ^{-1}\left(\frac{\omega \eta(T_s)}{\mu}\right)-\\
\tan ^{-1}\left(\frac{\omega \eta_m}{\mu}\right)\huge]
\end{split}
\label{eq:integrate-q}
\end{equation}
Assuming that the lithosphere is thin enough that the tidal deformation exactly matches the tidal geopotential, $\overline{\dot{\varepsilon}_{i j}^{2}}$ for planet with eccentricity $e$ is approximately $15\omega^2\gamma^2 e^2/4$ \citep{Ojakangas-Stevenson-1989:thermal}, where $\gamma=(a^3\omega^2)/(Gm_p)$ characterizes the relative magnitude of tidal forcing and the planet's self-gravity. 
Given that the bulk of heat production will take place in the lithosphere, 
we can obtain the heat flux at the top of the lithosphere $\mathcal{F}_s$ from Equation \ref{eq:integrate-q} and Equation \ref{eq:heat-flux-surf-bottom}:
\begin{equation}
\begin{split}
    \mathcal{F}_s&=-\kappa_0\left.\left( \frac{d \ln T}{d z}\right)\right|_{\mathrm{surf}}
    \\&= \sqrt{ \frac{15\kappa_0\mu\omega\gamma^2 e^2}{ l}\left[\tan ^{-1}\left(\frac{\omega \eta(T_s)}{\mu}\right)-\tan ^{-1}\left(\frac{\omega \eta_m}{\mu}\right)\right]}.
    \label{eq:heat-flux-surface}
    \end{split}
\end{equation}
This heat flux will be emitted to the space by radiation, heating the surface of the planet in the process. Assuming the surface radiates as a blackbody (Equation \ref{eq:radiation}), the surface temperature will be
\begin{equation}
\begin{split}
    T_s=\left(\frac{\mathcal{F}_s}{\sigma}\right)^{1/4}\approx \Bigg(&\frac{240\kappa_0\mu\omega M_*^2 e^2}{ m_p^2\sigma^2l}\bigg[\tan ^{-1}\left(\frac{\omega \eta(T_s)}{\mu}\right)\\
    &-\tan^{-1}\left(\frac{\omega \eta_m}{\mu}\right)\bigg] \Bigg)^{1/8},
\end{split}
\label{eq:Ts}
\end{equation}
when $\gamma=4M_*/m_p$.




In order to be able to produce any significant amount of heat, somewhere in the ice shell, the Maxwell time needs to be comparable to the orbital period. Since the Maxwell time at the water-ice interface is $\sim 0.4$~days, which is likely shorter than the orbital period, the aforementioned condition turns into 
\begin{equation}
   \omega \eta(T_s)/\mu>1
   \label{eq:Ts-mid}
\end{equation}
This sets an upper bound on surface temperature, below which dissipation given by Eq.\ref{eq:integrate-q} can be achieved. With $\omega\sim 10^{-5}$~s$^{-1}$, $\eta(T)=\eta_m(T_m/T)^l$, $\eta_m=10^{14}$~Pa$\cdot$s, and $\mu=3\times 10^{9}$~Pa, the above requirement leads to $T_s\sim T_m/3^{\frac{1}{14}}\sim T_m/1.08=252$~K (here we choose $l=14$ following \citet{Ojakangas-Stevenson-1989:thermal}). If surface temperature is lower than this threshold, which is very likely true, the planetary tidal quality factor $Q_p \sim l/2=7$ \citep{Ojakangas-Stevenson-1989:thermal}.
Otherwise, the dissipation rate will be lower than our estimate, yielding a higher quality factor. 

To simplify the calculation in the analytical treatment of this work, we assume $Q_p = 7$ for planets with surface temperatures below this cutoff temperature of 252 K. For planets above this cutoff temperature, we assume $Q_p = 100$, the standard value assumed for a stagnant lid Earth with no surface ice layer. 
Future numerical analysis will consider this problem of the time varying $Q_p$ with a higher resolution.







\subsection{Ocean Evaporation and Runaway Greenhouse}
\label{sec:greenhouse}
\begin{figure}
\centering
 \includegraphics[width=0.5\textwidth]{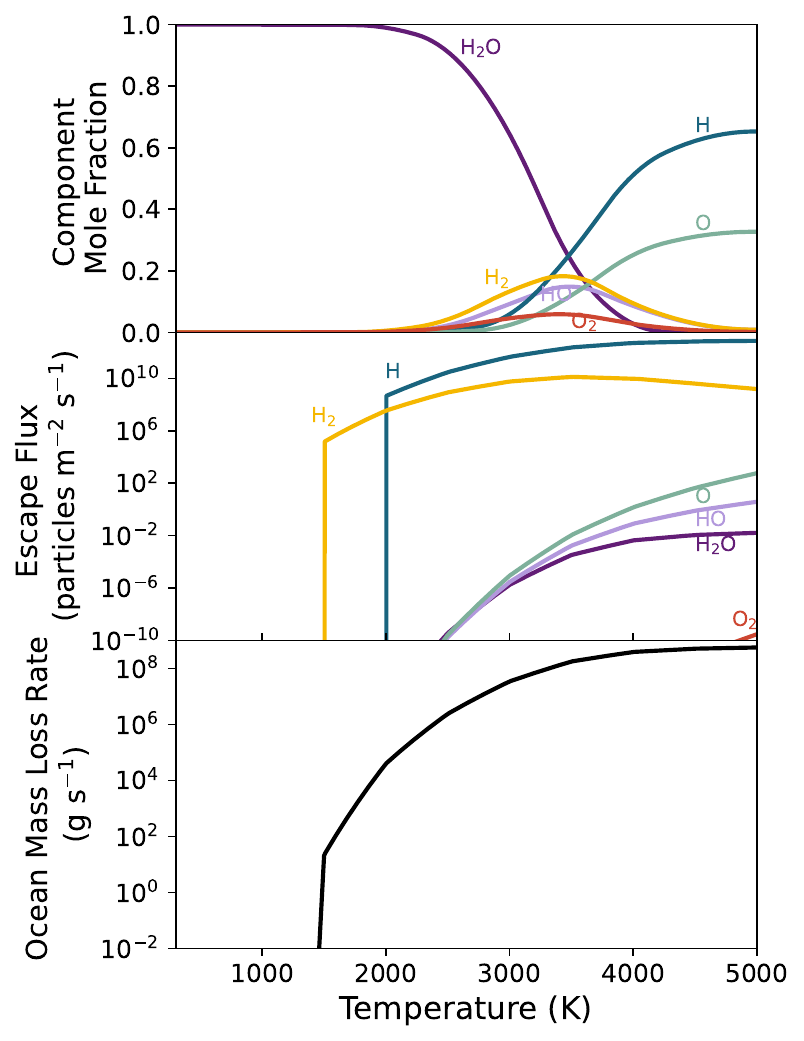}
 \caption{(Top panel) Molar fractions of different constituents of water vapor at a pressure of 1 atm, showcasing the relative abundances of each species \citep[data from][]{197959}. (Middle panel) Jeans escape rates, given in particles/m$^2$/s, for each constituent of water vapor on an Earth-like planet at varying temperatures. The harsh left edges in $H$ and $H_2$ particle escape rates are due to the lack of dissociated water vapor products at lower temperatures. (Bottom panel) The ocean mass loss rate in grams per second for an Earth-like planet at various temperatures. The calculation assumes that the planetary atmosphere is made entirely of water vapor and its dissociation products, and that no other sources of hydrogen or oxygen are available. The ocean mass loss rate only counts particles lost from the planet; otherwise, it is possible they could re-condense later at lower temperatures.  }
    \label{fig:watevapor}
\end{figure}

In Section \ref{sec:hothothot}, we explored mass loss driven by XUV radiation. This form of mass loss takes place when the XUV radiation directly heats the planet's upper atmosphere, disassociates water vapor molecules, and initiates a hydrodynamic flow that causes atmospheric particles to escape into space. However, when the planet's temperature is primarily set by tidal heating, the situation differs. Tidal heating generates heat within the planetary interior rather than in the upper atmosphere, and this interior heat does not directly cause the atmospheric particles to enter a hydrodynamic flow. Instead, molecules in the upper atmosphere can be lost via Jeans escape, a process where thermal motion allows particles to reach escape velocity and overcome the planet's gravitational pull. Similarly, without incident XUV radiation to dissociate water vapor molecules \citep{Kasting1983}, the atmosphere must reach a much higher temperature (2000-3000 K) for water vapor to dissociate into hydrogen and helium (see top panel of Figure \ref{fig:watevapor}).

The first thing that must happen is that some amount of the liquid ocean must evaporate into the atmosphere. This process is dependent on the planet's surface temperature and the vapor pressure of water at that temperature, as described by the water vapor pressure relation \citep{Kasting1988}. As the surface temperature increases, a higher fraction of the ocean's water transitions into its gaseous form, enriching the planet's atmosphere with water vapor. Once the water vapor has entered the atmosphere, its fate is determined by a combination of factors including temperature-dependent dissociation into hydrogen and oxygen, as well as potential escape mechanisms such as Jeans escape. Therefore, the initial evaporation of the ocean serves as the gateway process, setting the stage for subsequent thermal and dynamical interactions that could lead to the loss or retention of the planet's initial water mass.

The atmospheric composition is treated as a heterogeneous mixture comprising water vapor, hydrogen, and oxygen. The relative concentrations of these constituents are governed by the atmospheric temperature, with higher temperatures promoting dissociation and consequently, a higher prevalence of elemental hydrogen and oxygen (see Figure \ref{fig:watevapor}).

The fractional Jeans thermal escape flux $\Phi_J$ is set by the relative speeds of the particle velocities and escape velocity: 
\begin{equation}
    \Phi_J = \frac{n_{exo} v_{0}}{2\sqrt{\pi}} \left(\frac{v_{e}^2}{v_0^2} + 1 \right)e^{-v_{e}^2/v_0^2},
\end{equation}
where $n_{exo}$ is the number density of the escaping component at the exobase, $v_{e}$ is the escape velocity, defined as
\begin{equation}
    v_e = \sqrt{2 G m_p / r_p},
\end{equation}
and the velocity of escaping particles $v_0$ is
\begin{equation}
    v_0 = \sqrt{2 k_B T / m_{H}}.
\end{equation}
The exobase is defined as the lower limit of the exosphere (where particle collisions are no longer frequent enough to maintain a Maxwellian velocity distribution). The number density at the exobase can thus be defined as: 
\begin{equation}
    n_{exo} = (m_H * \mu) * g / (a k_B * T) 
\end{equation}
where $a$ is the particle cross section and $g$ the gravitational acceleration.

The temperature here will be higher than the equilibrium temperature of the planet, although the dominant molecular component of the atmosphere will affect the exact heat (for example, atmospheres dominated by CO$_2$ will produce cooler exospheric temperatures compared to those dominated by hydrogen and helium; \citealt{Konatham2020}). In this work, for simplicity, we assume an isothermal atmosphere where the exobase temperature is equal to the surface equilibrium temperature.

The rate of particle loss $\dot{n}$ from the planet due to Jeans escape for a species with molecular mass $\mu_g$ is therefore given by:
\begin{equation}
\begin{split}
    \dot{n} &= 4 \pi r_p^2 \Phi_J \\
    &= \frac{4 \pi r_p^2 v_{0} m_H  \mu g}{ 2\sqrt{\pi} a k_B T }  \left(\frac{v_{e}^2}{v_0^2} + 1 \right)e^{-v_{e}^2/v_0^2}
    \end{split}
\end{equation}
For the bulk of XUV flux rates under consideration, this mass loss rate will be less than the mass loss rate given in Section \ref{sec:hothothot}. However, heat provided by tidal migration may fractionally evaporate a planet's atmosphere even if particle mass loss rates are slow. 








%
%
%
%
%
%
%
%

\section{Ocean Retention by Initial Planet Parameters}
\label{sec:results}
Simultaneously with the migration process, which creates heat in the interior of the planet, the migrating planet will also be irradiated by the white dwarf. We can adapt Equation \ref{eq:radiation} to account for the eccentric orbit, following \citet{Adams2006ecc} and \citet{Gallo2024}:
\begin{equation}
F_p(t) = \dfrac{F_p(t)}{\sqrt{1 - e^2}} = \frac{(1-A) L_{\star}(t)}{4 \pi a(t)^2 \sqrt{1 - e^2}}.
\label{eq:radiation_updated}
\end{equation}
Unlike in the previous section, where orbits were considered to be roughly circular as the system evolved, the high eccentricities attained during planet migration mean that the planet experiences a larger time-averaged flux than the same planet would experience in a circular orbit with the same semi-major axis. 

To compute the final temperature on a migrating planet, we must combine the total heat contributed by input flux from the white dwarf (computed directly using Equation \ref{eq:radiation_updated}) with the portion of the heat input contributed by tidal heating that is available to heat the surface of the planet:
\begin{equation}
F_{surf} = F_{p}(t) + \frac{\dot{E_p}(t)}{4 \pi r_p^2}.
\label{eq
}
\end{equation}
For these large values of $\dot{E_p}(t)$, we also assume that tidal heating has evaporated the surface ocean into a steam atmosphere in a greenhouse state \citep{Kasting1988}. 
This time-varying expression is used to compute the surface temperature, $T_{s}$, of the planet. Because the planet scattering event that begins the tidal migration process and generates $\dot{E_p}(t)$ may occur at any white dwarf cooling age, the heat input will also vary accordingly.

The surface temperature, determined by both radiation and tidal heating, plays a critical role in setting the evaporation fraction of the ocean into the atmosphere. The temperature $T_{s}$ influences the rate of ocean mass loss, computed by summing the mass loss rate due to photoevaporation and the mass loss rate due to tidal heating.  
The evaporated fraction of the ocean, along with the surface temperature, dictates the mass loss due to photoevaporation and hydrodynamic outflow. These factors collectively determine the amount of surface water remaining on the planet. The latter requires much higher surface temperatures for outflows, so while tidal heating is effective at evaporating a surface ocean into the atmosphere, photoevaporation generally drives the bulk of permanent mass loss from the planet. 

The exact dynamics of ocean loss depend sensitively on a large quantity of initial conditions including planet mass, radius, ocean quantity, semi-major axis, timing of migration, white dwarf mass, etc. As a result, there is no universal answer for how much surface water will be retained during the processes outlined in this work.
However, we can present a few representative cases to illustrate the range of possibilities.

\begin{figure}
\centering
 \includegraphics[width=0.45\textwidth]{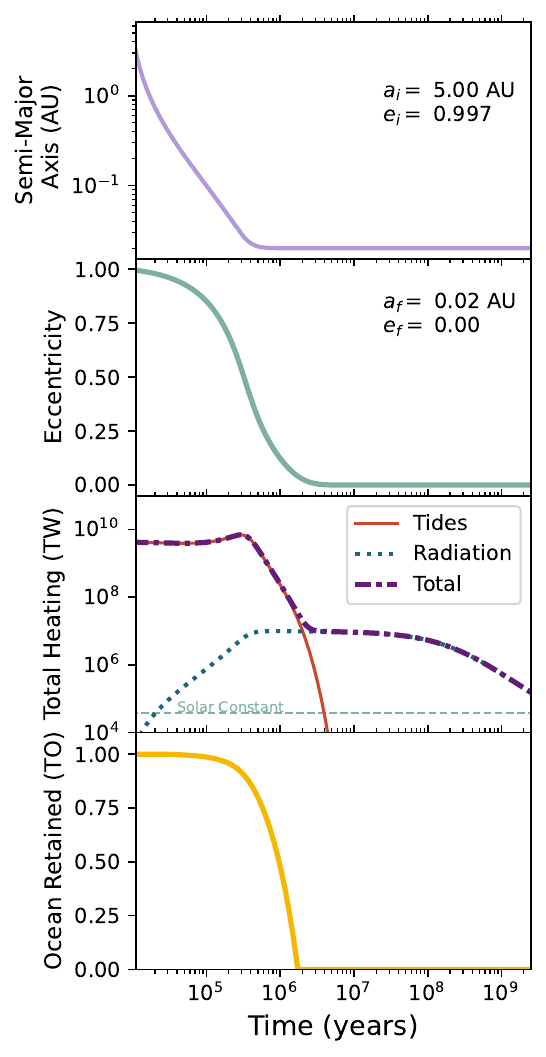}
 \caption{Evolution of planetary parameters and heating over time. The top two panels show the change in semi-major axis and eccentricity from initial values ($a_i$ = 5.00 AU, $e_i$ = 0.997) to final values ($a_f$ = 0.02 AU, $e_f$ = 0.00). The third panel presents the total heating (TW) from solar radiation and tidal forces. The bottom panel illustrates the fraction of oceans lost over time (in terrestrial oceans, TO) due to the combined effects of radiation and tidal heating. For this migration process, which begins with a scattering event that occurs shortly after the formation of the white dwarf, no surface water remains.  }
    \label{fig:Fig1}
\end{figure}

To start, in Figure \ref{fig:Fig1}, we show the ocean loss (bottom panel) for a migrating planet whose scattering event takes place shortly after the formation of the white dwarf. For this planet, the tidal heating is sufficient to fully evaporate the ocean into the atmosphere of the planet on relatively short timescales. Since the migration moving the planet inwards to hotter orbits takes place early when the white dwarf is still hot and bright in the XUV, once this water is in the atmosphere as water vapor, it is fully lost fairly quickly (approximately 1 Myr after the scattering event).

\subsection{Timing of Planet Scattering Events}
For a planet like the ones we describe above to be placed into a habitable orbit via tidal migration, tidal migration must be precipitated by a scattering event which increases the planetary eccentricity to near white dwarf-crossing values. Numerical simulations of planet stability in the white dwarf phase show that due to the chaotic nature of planet interactions post-system-reshuffling that occurs at the start of the white dwarf phase, this scattering event may occur immediately after white dwarf formation, but it is more likely that it will take a while for it to onset \citep{Veras2013}. 

It is possible that such a scattering event may be delayed by several Gyr. The exact timing depends on the dynamics of the system and the Lyapunov time, but this variability in the onset of the process of tidal heating driven by planet migration introduces additional complexity in predicting the thermal and hydrodynamic evolution of the planet’s surface water.

\begin{figure*}
\centering
 \includegraphics[width=\textwidth]{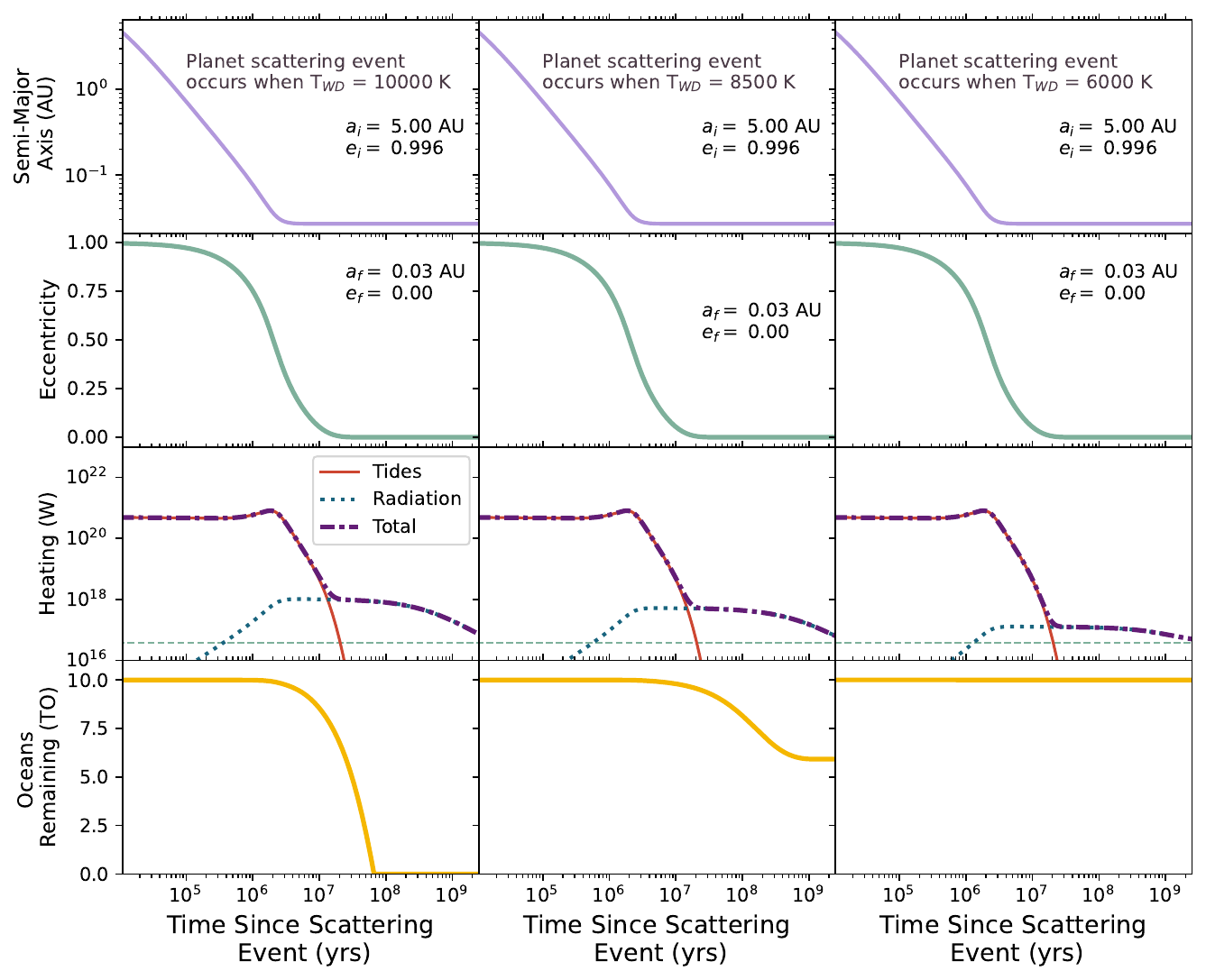}
 \caption{Evolution of planetary parameters following a scattering event that excites a planet to large eccentricity. The top row shows the semi-major axis evolution of a planet after the scattering event, which occurs when the white dwarf's temperature is (first column) 10000 K, (second column) 8500 K, and (third column) 6000 K. The second row shows eccentricity evolution computed. The third row displays the heating rates from tides and radiation over time, along with the total heating experienced by the planet. The bottom panel illustrates the fraction of oceans remaining as a function of time since the scattering event, highlighting the significant impact of these events on the long-term habitability of the planet. The initial and final semi-major axes ($a_f$ and $a_i$) and eccentricities ($e_f$ and $e_i$) are labeled for each case. For scattering events that occur later (once the white dwarf has already cooled), ocean retention is easier as both the XUV luminosity fraction and the total white dwarf luminosity are lower.  }
    \label{fig:fig12frame}
\end{figure*}

In Figure \ref{fig:fig12frame}, we show an analogous plot to Figure \ref{fig:Fig1} for a set of planet initial conditions and for three cases (columns): three different times for the scattering event, which take place when the white dwarf is 10000 K (roughly 0.5 Gyr old), 8500 K (roughly 0.8 Gyr old), and 6000 K (roughly 2 Gyr old). 
For illustrative purposes, we increase the initial ocean budget of the planet to 10 TO to demonstrate the effect of delaying the scattering event that places the planet into a high-eccentricity orbit and stars the process of tidal migration. 

The XUV radiation of the white dwarf, which is what primarily drives atmospheric escape of the evaporated ocean, decreases as the white dwarf ages for two reasons: first, the total luminosity of the white dwarf decreases as it cools; second, the XUV fraction of its emitted luminosity also decreases as its temperature decreases.
Because of this, the effect of delaying the scattering event (which delays the time when the planet reaches short orbital periods until after the white dwarf has cooled and its XUV output has decreased) is to allow more of the ocean to survive.

\begin{figure}
\centering
 \includegraphics[width=0.5\textwidth]{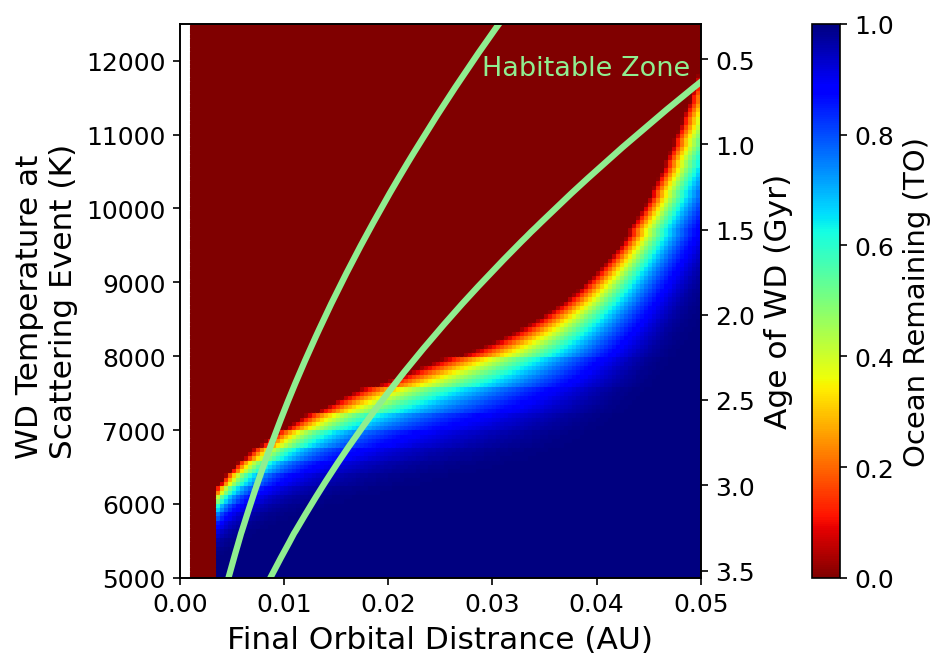}
 \caption{A parameter space plot showing ocean survival plotted by planet final semi-major axis and the time since white dwarf formation at which the scattering event occurred. This plot was constructed for an Earth-like planet with an initial semi-major axis of 5 AU, varying initial eccentricity, and an initial ocean mass of 1 TO. The location of the habitable zone, defined by where water could be liquid on the surface of an Earth-like planet, is overlaid.  }
    \label{fig:final24}
\end{figure}

In Figure \ref{fig:final24}, we show a parameter space plot demonstrating how the timing of the scattering event affects the planet's ocean retention, overlaid with the approximate location of the habitable zone (defined by the orbital locations where a planet with Earth-like albedo and atmospheric composition could host liquid water). This parameter space was constructed for a planet that started its tidal migration from a distance of 5 AU and had a total surface water at that time of 1 TO (which means that it would have necessarily had a larger amount of surface water on the main sequence). In general, planets are less likely to retain oceans if their final orbital distance is smaller, and planets are less likely to retain oceans if the scattering event occurs earlier, all the white dwarf is hotter. Since the location of the habitable zone ranges from 0.01 to 0.1 AU, this specific planet does allow ocean retention in the approximate habitable zone so long as the migration process begins fairly late (3 Gyr). At a white dwarf cooling time of 3 Gyr, the habitable zone location is estimated to be centered on 0.01 - 0.02 AU \citep{Agol2011}. 



\section{Discussion}
\label{sec:discussion}

In this paper, we have investigated the retention of surface water on first-generation planets orbiting white dwarfs. 
{While the planets considered in this work were likely not habitable during their host star's main sequence phase, their potential for habitability may emerge after dynamical interactions bring them into the habitable zone of the white dwarf.}
Our work considered the various stages of stellar evolution, including the red giant and AGB phases, and their impact on planetary orbits and surface conditions. We modeled the processes of ocean evaporation due to increased stellar luminosity and subsequent photoevaporation driven by X-ray and extreme ultraviolet (XUV) radiation from the newly formed white dwarf. 
Additionally, we explored the effects of tidal heating during planetary migration to ocean evaporation and subsequent atmospheric mass loss. By combining these models, we assessed the potential for planets to retain their oceans under different initial conditions and evolutionary scenarios. Below, we summarize our main conclusions:
\begin{enumerate}
    \item \textbf{Water Retention on Planets}: Earth-sized or even Mars-sized planets with initial water reserves are unlikely to retain their oceans unless they possess significantly larger initial amounts of surface water compared to Earth.
    \item \textbf{Initial Conditions for Water Retention}: Planets capable of maintaining their water reserves must either begin with vast initial quantities of water or originate in very distant orbits (greater than 5-6 AU for a planet with 1 TO of initial water) from their stars, or both.
    \item \textbf{Impact of Tidal Heating and XUV Radiation}: While tidal heating can contribute to ocean evaporation and greenhouse effect, it is generally insufficient to completely strip the atmosphere. XUV radiation is necessary to drive the full loss of surface water to space.
    \item \textbf{Timing of Scattering Events}: As a result of the previous point, delaying the scattering event can enhance ocean survival prospects. 
\end{enumerate}

\subsection{Worst-Case Scenario and Considerations}
The calculations in this paper represent a worst-case scenario in terms of water availability, as we assume that water can only be lost from the migrating planet. 
Water resupply via late volatile delivery \citep{Hartogh2011, Alexander2012, Ciesla2015, Sanchez2018, Wyatt2020, Seligman2022} from comets could enable planets that our model suggests would not support a surface ocean to retain one.

Because one of the main motivations for studying terrestrial planets around white dwarfs is the search for biosignatures, it is important to consider how the atmospheres of these planets may be different compared to those of planets around main sequence stars because of the evolutionary processes that they endure as their host star evolves off the main sequence. A potential risk for planets undergoing significant ocean evaporation and atmospheric mass loss, as highlighted by \citet{Luger2015}, is the buildup of atmospheric O$_2$. Evaporating oceans and losing atmospheric hydrogen through XUV-driven mass loss without mechanisms to sequester O$_2$ could lead to an atmosphere rich in O$_2$, creating a false positive for life detection. 

Another concern for habitability is a persistent greenhouse effect. An atmospheric greenhouse effect (supplied by an atmosphere with significant quantities of steam) could either warm planets that would otherwise be too cold to be habitable, or make apparently habitable planets too warm to host life. The strength of this effect will depend on exact atmospheric composition, resulting in the habitability limits and ocean retention values presented in Figure \ref{fig:final24} not being globally applicable to all planet types.

\subsection{Future Work}
There are several avenues for further research that could enhance our knowledge of water retention on planets orbiting white dwarfs. 
Future work will benefit from more sophisticated models and targeted observational campaigns to address the complexities of planetary habitability in these unique environments, and determine if there are any candidate planets upon which the mechanisms discussed in this paper could be relevant. 

\subsubsection{Improved Modeling}
Future research should employ more complex numerical models to assess the full habitability of planets, considering additional processes beyond our simple analytical models. While maintaining oceans on planets with Earth-like amounts of surface water or short orbital periods (a few AU) is challenging, planets with significantly larger initial water budgets might retain their oceans. However, the presence of surface water alone does not guarantee habitability, as the full implications of a greenhouse effect and other factors must be considered in future work. 

On this work, we have only considered Earth-like planets (defined as those with a radius of 1 $R_{\oplus}$ and mass of 1 $M_{\oplus}$), and we have not considered how the surface distribution of water affects the results. Future works could benefit from exploring the effect of planet type (icy moon sized vs super-Earth sized) and surface water distribution (water distributed evenly across the surface or in ice caps at the poles) on ocean retention. 

Additionally, the results in this study are based on stellar evolution tracks for a 1 $M_{\odot}$ star. The duration of the TP-AGB phase, where significant ocean mass loss occurs, is highly dependent on the star's mass \citep{Marigo2015}. Consequently, similar evolution around larger or smaller progenitor stars could yield significantly different results.

\subsubsection{Observational Prospects}
One of the key questions for future exploration in this topic is whether progenitor planets capable of evolving into terrestrial habitable zone planets around white dwarfs exist. Thus far, no intact terrestrial planets around white dwarfs have been detected. To find these planets will require a combination of continued transit surveys that can detect short-duration transit events with a bit of luck (as the transit probabilities for planets around white dwarfs are very low), or potentially non-transit-based surveys as proposed by \citet{Limbach2022}. 

It is also worth investigating the potential of  moons as potential habitable objects around white dwarfs \citep[ex:][]{Limbach2023}. While we have thus far only our own Solar System as an example here, moons formed around gas giants past a system's ice line may possess significant ice or water reserves that could survive during the star's post-main-sequence evolution. These moons could be liberated from their host planets during the post-main-sequence evolution or planet migration \citep{Spalding2016}, and their potential for habitability could be a fruitful avenue for future study.

\section{Conclusion}
\label{sec: conclude}
In this work, we explored the processes affecting the retention of surface oceans on first-generation planets orbiting white dwarfs. Our models considered the significant challenges posed by stellar evolution, including the red giant and asymptotic giant branch phases, followed by the intense XUV radiation emitted by the newly formed white dwarf. We found that planets starting with large initial reservoirs of water and located at considerable distances from their host stars are more likely to retain their oceans. Moreover, the timing of planetary migration and the onset of tidal heating plays a crucial role in determining the fate of these water reserves. If a scattering event that initiates planetary migration occurs after the white dwarf has sufficiently cooled, the feasibility of ocean retention increases due to reduced XUV radiation.
While our results highlight the difficulties in maintaining surface water on planets transitioning to the habitable zones of white dwarfs, they also suggest that under specific conditions, such retention is possible. Planets with substantial initial water masses, or those that migrate inwards late in the white dwarf's cooling phase, stand a better chance of sustaining their oceans.


\acknowledgments
We thank Wanying Kang and Ylva Goetberg for significant assistance and useful suggestions. We also thank Melinda Soares-Furtado, Elena Gallo, Fred Adams, Darryl Seligman and Konstantin Batygin for useful conversations.  
{We thank the referee for their comments and suggestions.}

\vspace{5mm}


\bibliography{refs}{}
\bibliographystyle{aasjournal}



\end{document}